\documentclass{aa}  
\usepackage{graphicx}
\begin{document}
   \title{Chemical evolution of the Small Magellanic Cloud based on planetary nebulae
        \thanks{Tables 2, 3, and 7 are only available in electronic form at the CDS
         via anonymous ftp to cdsarc.u-strasbg.fr (130.79.128.5) or via 
         http://cdsweb.u-strasbg.fr/cgi-bin/qcat?J/A+A/}}

   \author{T. P. Idiart
          \inst{1},
          W. J. Maciel
          \inst{1},
          \and
          R. D. D. Costa\inst{1}
          }

   \offprints{T. P. Idiart}

   \institute{Instituto de Astronomia, Geof\'\i sica e Ci\^encias Atmosf\'ericas
    (IAG/USP), Universidade de S\~ao Paulo, Rua do Mat\~ao 1226, 05508-900 S\~ao
    Paulo SP, Brazil\\
              \email{thais@astro.iag.usp.br, maciel@astro.iag.usp.br, roberto@astro.iag.usp.br } 
              }

   \date{Received ..., ...; accepted ..., ...}

% \abstract{}{}{}{}{} 
% 5 {} token are mandatory
 
  \abstract
  % context heading (optional)
   {}
   % aims heading (mandatory)
   {We investigate the chemical evolution of the Small Magellanic Cloud
   (SMC) based on abundance data of planetary nebulae (PNe). The main 
   goal is to investigate the time evolution of the oxygen abundance in
   this galaxy by deriving an age-metallicity relation. Such a relation
   is of fundamental importance as an observational constraint of chemical 
   evolution models of  the SMC.}
  % methods heading (mandatory)
   {We have used high quality PNe data in order to derive the properties of the
   progenitor stars, so that the stellar ages could be estimated. We collected 
   a  large number of measured spectral fluxes for each nebula, and derived 
   accurate physical parameters and nebular abundances. New spectral data for 
   a sample of SMC PNe obtained between 1999 and 2002 are also presented. These 
   data are used together with data available in the literature to improve the 
   accuracy of the fluxes for each spectral line.}
  % results heading (mandatory)
   {We obtained accurate chemical abundances for PNe in the Small Magellanic Cloud,
   which can be useful as tools in the study of the chemical evolution of this 
   galaxy and of Local Group galaxies. We present the resulting oxygen versus age 
   diagram and a similar relation involving the [Fe/H] metallicity based on a 
   correlation with stellar data. We discuss the implications of the derived
   age-metallicity relation for the SMC formation, in particular by suggesting
   a star formation burst in the last 2--3 Gyr.}
  % conclusions heading (optional) 
   {}

   \keywords{abundances -- Small Magellanic Cloud
     -- planetary nebulae -- chemical evolution
               }
   \authorrunning{T. P. Idiart et al.}
   \titlerunning{Chemical evolution of the SMC}

   \maketitle
%
%_________________________________________________________________________________

\section{Introduction}

A comprehensive study of the chemical enrichment in a given stellar system involves 
the determination of accurate abundances and the building of several chemical diagrams 
showing the evolution of the abundances of different elements and their variation with 
time. In particular, diagrams of the abundances as a function of age  are fundamental 
in order to decrease the number of possible solutions of chemical evolution models, 
working as a strong constraint to these models. The understanding of the evolutionary 
process of the Magellanic Clouds, and of the Small Magellanic Cloud (SMC) in particular, is 
important in many aspects:  the metallicity of the SMC is closer to that of a primordial 
galaxy than the Large Magellanic Cloud (LMC) or the Galaxy, its distance is well known 
and the local reddening is considerably lower than that of the Galactic disk.  

The star formation histories of the LMC and SMC seem to present different patterns
from each other, and are in many respects still controversial (cf. Olszewski et al. 
\cite{olszewski}). Bertelli et al. (\cite{bertelli}) pointed out that the LMC 
experienced an episode of star formation around 3-5 Gyr ago, while the star formation 
history of the SMC would indicate a constant formation rate during the last 2-12 Gyr 
(Dolphin et al. \cite{dolphin}). On the other hand, Piatti et al. (\cite{piatti2}) 
studied a sample of clusters in the SMC and concluded that there is a peak in their 
age distribution at 2.5 Gyr, which corresponds to a very close encounter between the 
LMC and the SMC according to dynamical models, in agreement with the results from the 
bursting model by Pagel and Tautvai\v siene (\cite{pagel}), adopting a burst that 
occurred 3 Gyr ago. More recent work on the SMC, especially on the basis of the age 
distribution of stellar clusters, is consistent with a star formation burst in the 
last few Gyr, as we will discuss in more detail in Section~6.

In this work, we obtain accurate chemical abundances for planetary nebulae in the SMC
and use these results to study the chemical evolution of this galaxy. To do this, 
high quality PNe data are needed, especially to derive the properties of PNe 
progenitors to estimate their ages. Our first goal was to collect a large number of 
measured spectral fluxes for each SMC PNe, in order to derive accurate physical parameters 
and abundances. We also present new spectral data for a sample of SMC PNe obtained 
by our group between 1999 and 2002. These data are combined with data available in the 
literature to improve the accuracy of the fluxes for each PNe spectral line. Finally, we 
present an age-metallicity relationship in the form of oxygen abundances relative 
to the sun as a function of age and [Fe/H]--age diagrams, and discuss their implications 
for SMC formation.  

%_________________________________________________________________________________

\section{The sample}

\subsection{Observations and data reduction}

Observations were performed using telescopes at ESO – La Silla (1.52 m) and at the
Laborat\'orio Nacional de Astrof\'\i sica (LNA), Brazil (1.60 m). Weather 
conditions are normally different in both observatories: at ESO/La Silla the 
average seeing is typically smaller than 1", and for measurements at this site 
we used slit width of 1". On the other hand, at LNA we adopted 1.5" slit width 
due to the poorer seeing conditions. All measurements were performed with airmass 
smaller than $X$ =1.5 in order to keep the spectrophotometric accuracy of the 
results. In both observatories, long east-west slits were used in all cases. 
Also, in both observatories,  Boller \& Chivens Cassegrain spectrographs were used; 
at ESO with CCD and grating allowing a reciprocal dispersion of about 2.5 \AA/pixel, 
and at LNA with a CCD and grating with smaller dispersion, namely, 4.4 \AA/pixel. 
This sample includes 36 objects, and the log of the observations is given 
in Table~\ref{logobs}. In this table we also included the exposures times
$t_{ex}$ in seconds and the extinction coefficient $c_{ext}$ derived from the 
H$\alpha$/H$\beta$ ratios assuming Case B (Osterbrock \cite{osterbrock}) and 
adopting the extinction law by Cardelli et al. (\cite{cardelli}). The PNe
marked with an asterisk have been included in Costa et al. (\cite{costa}),
but for 8 of these objects new observations have been made.

%
% TABLE 1: observations log *** IN THE PAPER *** -------------------------------------------
%
   \begin{table*}
      \caption{Log of the observations.}
      \label{logobs}
%      \centering
      \begin{tabular}{llcccllccc}
      \hline\hline
          Object    & Date  &  Site  & $t_{ex} (s) $ & $c_{ext}$ & Object    & Date  &  Site  
    & $t_{ex} (s)$ & $c_{ext}$ \\
      \hline
       SMP 1* & 1999 Aug 19 & ESO & 3600 & 0.18 &\ SMP 16* & 1999 Dec 29 & ESO & 2400 & 0.03 \\
              & 2002 Set 25 & ESO & 3600 & 0.23 &\ SMP 17* & 1999 Aug 17 & ESO & 1800 & 0.64 \\	   
       SMP 2* & 1999 Aug 19 & ESO & 2400 & 0.18 &\ SMP 18* & 1999 Aug 17 & ESO & 2400 & 0.28 \\   
       SMP 3* & 1999 Aug 19 & ESO & 2100 & 0.07 &\         & 2002 Oct 11 & ESO & 3600 & 0.37 \\
       SMP 4* & 1999 Aug 20 & ESO & 2400 & 0.00 &\ SMP 19* & 1999 Aug 18 & ESO & 3600 & 0.45 \\
              & 2002 Oct 12 & ESO & 3000 & 0,00 &\ SMP 20  & 2002 Oct 11 & ESO & 2400 & 0.00 \\
       SMP 5* & 1999 Jul 18 & LNA & 2400 & 0.67 &\ SMP 21* & 1999 Dec 28 & ESO & 2400 & 0.26 \\
       SMP 6* & 1999 Jul 18 & LNA & 2400 & 0.42 &\ SMP 22* & 1999 Aug 17 & ESO & 1800 & 0.39 \\
              & 2002 Set 25 & ESO & 5400 & 0.45 &\ SMP 23* & 1999 Aug 20 & ESO & 2400 & 0.00 \\
              & 2002 Oct 13 & ESO & 2400 & 0.61 &\ SMP 24  & 2002 Oct 10 & ESO & 3000 & 0.03 \\
       SMP 7* & 1999 Aug 15 & ESO & 3600 & 0.15 &\ SMP 25* & 1999 Dec 28 & ESO & 3000 & 0.00 \\
       SMP 8* & 1999 Aug 20 & ESO & 2400 & 0.03 &\ SMP 26  & 2002 Set 27 & ESO & 3600 & 0.00 \\
              & 2002 Set 27 & ESO & 2400 & 0.55 &\ SMP 27  & 2002 Set 27 & ESO & 2400 & 0.00 \\
       SMP 9* & 1999 Jul 20 & LNA & 2400 & 0.95 &\ SMP 28  & 2002 Oct 10 & ESO & 3600 & 0.00 \\
              & 1999 Dec 28 & ESO & 3000 & 0.92 &\ MGPN 1  & 2002 Oct 09 & ESO & 3600 & 2.38 \\
       SMP 10*& 1999 Jul 20 & LNA & 2400 & 0.90 &\ MGPN 2  & 2002 Oct 10 & ESO & 3600 & 2.87 \\
              & 2002 Set 28 & ESO & 3600 & 0.07 &\ MGPN 3  & 2002 Oct 11 & ESO & 3600 & 0.16 \\
       SMP 11*& 1999 Dec 26 & ESO & 2400 & 0.59 &\ MGPN 5  & 2002 Oct 12 & ESO & 3600 & 0.00 \\
              & 2002 Oct 08 & ESO & 3600 & 0.62 &\ MGPN 7  & 2002 Oct 13 & ESO & 3600 & 0.99 \\
       SMP 12*& 1999 Aug 16 & ESO & 3600 & 0.00 &\ MGPN 11 & 2002 Oct 09 & ESO & 3600 & 1.72 \\
       SMP 13*& 1999 Dec 26 & ESO & 2400 & 0.05 &\ MGPN 13 & 2002 Set 27 & ESO & 2400 & 0.39 \\
              & 2002 Set 28 & ESO & 3600 & 0.64 &\ [M95] 9*& 1999 Aug 15 & ESO & 3600 & 0.24 \\
       SMP 14*& 1999 Dec 27 & ESO & 2400 & 0.05 &          &             &     &       \\
       SMP 15 & 1999 Aug 16 & ESO & 3600 & 0.26 &          &             &     &       \\
              & 2002 Oct 10 & ESO & 3000 & 0.00 &          &             &     &       \\
      \hline
      \end{tabular}
   \end{table*}

\begin{figure}
   \centering
   \includegraphics[angle=-90, width = 8cm]{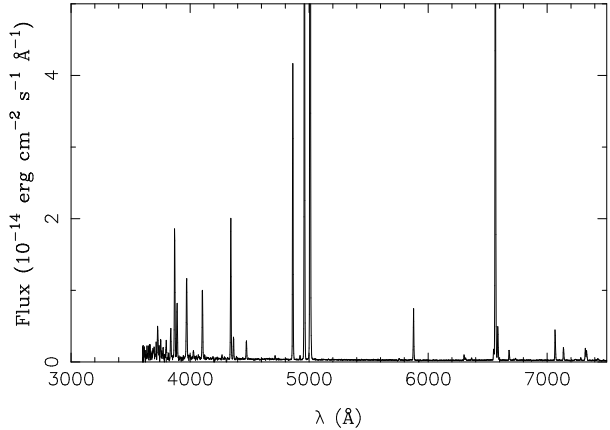}
   \caption{An example of a spectrum of the planetary nebula SMP 15 in the SMC.}
   \label{spectrum}
   \end{figure}

Image reduction and analysis were performed using the IRAF package, including the classical 
procedure to reduce long slit spectra: bias, dark and flat-field corrections, spectral 
profile extraction, wavelength and flux calibrations. Atmospheric extinction was corrected 
using mean coefficients for each observatory, and flux calibration was secured by the 
observation of standard stars (at least three) every night. Emission line fluxes were 
calculated assuming Gaussian profiles, and a Gaussian de-blending routine was used when 
necessary. Fig.~\ref{spectrum} presents a sample spectrum for the planetary nebula
SMP 15 in the SMC, taken in October, 2002.

%
%--------------------------------------------------------------------------------
%--------------------------------------------------------------------------------

The PNe sample given in Table~\ref{logobs} was observed in order to improve the measured 
line fluxes. In Table~2 (available in electronic form at the CDS), we report the new 
measured fluxes with errors including those from Costa et al. (\cite{costa}) for the lines 
used in this work. These fluxes were corrected from reddening and are presented in the 
H$\beta$ = 100 scale. Typical errors at the end of the spectrophotometric calibration 
process depend mainly on the derived sensitivity function for each night, which is related 
to distinct factors like the S/N for each line, the adopted atmospheric extinction
coefficient and photometric quality of the night. Uncertainties in the
individual measurements of each line are estimated as about 4\% for lines stronger
than 100 (in the H$\beta$=100 scale), 10\% for lines between 10 and 100, 15\% for 
lines between 1 and 10, and 32\% for lines weaker than 1 in the same scale.
The errors quoted in Table~2 are the resulting dispersions from the average on the
individual measurements. In this table, the first column gives the
wavelength (in \AA) and the ion responsible for the transition, while the remaining
columns give the measured fluxes in the H$\beta$ = 100 scale, followed by the estimated 
uncertainty and the number of measurements taken. In this table, as in all remaining tables 
of this paper, no errors are given when only one measurement was available.

%
% TABLE 2: New measured fluxes (relative to H beta) *** CDS *** ------------------------------------
%

\subsection{Additional objects}

In order to obtain a more significant sample, we have also taken into account some SMC 
nebulae from the literature, as well as flux measurements of the nebulae given in 
Table~\ref{logobs} by different sources. The new objects are SMP 32, SMP 34, MGPN 6, 
MGPN8, MGPN 9, MGPN 10, MGPN 12, Ma01, Ma02, and [M95] 8. The total sample includes then 
46 objects, and is shown in Table~3. In this table, which is also available at the CDS, 
the final flux data used to derive the physical parameters of the nebulae are presented 
in the same form as in Table~2. The line fluxes given in this table are simple averages of 
our data and other data compiled from the literature, taken from Webster (\cite{webster}), 
Osmer (\cite{osmer}), Dufour \& Killen (\cite{dufour}), Aller et al. (\cite{aller}), 
Monk et al. (\cite{monk}), Boroson \& Liebert (\cite{boronson}), Meatheringham \& Dopita 
(\cite{meatheringham1}, \cite{meatheringham2}), Vassiliadis et al. (\cite{vassiliadis}), 
Meyssonnier (\cite{meyssonier}), and Leisy \&  Dennefeld (\cite{leisy1}). Too discrepant 
flux values were discarded. 

%
% TABLE 3: Average fluxes (relative to H beta) *** CDS *** ------------------------------------
%

The CIII] $\lambda$1907,1909 fluxes as given in Table~3 were measured from IUE 
spectra, adopting an average extinction coefficient. This coefficient was taken as
a simple, unweighted average from several references, and is given in Table~\ref{extinction}, 
where the original references are shown. These references include Monk et al. (\cite{monk}),
Boroson \& Liebert (\cite{boronson}), Meatheringham et al. (\cite{meatheringham1},
\cite{meatheringham2}), Vassiliadis et al. (\cite{vassiliadis}), Meyssonier 
(\cite{meyssonier}), Costa et al. (\cite{costa}), Stanghellini et al. (\cite{stanghellini}),  
Leisy (\cite{leisy2}), and Leisy \& Dennefeld (\cite{leisy3}), apart from our own measurements.

%
% TABLE 4: average extinction coefficients *** IN THE PAPER *** --------------------------------
%

\setcounter{table}{3}
\begin{table*}
\caption{Average extinction coefficients and references.}
\label{extinction}
%      \centering
\begin{tabular}{llllll}
\hline\hline
Object & $\langle c_{ext} \rangle$ & References$^{*}$ & Object & $\langle c_{ext} \rangle$
 & References$^{*}$ \\
\hline
SMP 1  & $0.26\pm 0.07$ & 3,1,2,8,9      & SMP 24   & $0.03\pm 0.03$ & 1,2,9     \\
SMP 2  & $0.28\pm 0.09$ & 3,2,6,8        & SMP 25   & $0.19\pm 0.19$ & 1,2,6,9   \\
SMP 3  & $0.06\pm 0.10$ & 3,2,6,8        & SMP 26   & $0.31\pm 0.24$ & 1,2,6,9   \\
SMP 4  & $0.17\pm 0.25$ & 3,1,2          & SMP 27   & $0.03\pm 0.04$ & 3,2,6,8,9 \\
SMP 5  & $0.43\pm 0.28$ &  3,6,8         & SMP 28   & $0.15\pm 0.09$ & 3,2,6,8   \\
SMP 6  & $0.57\pm 0.15$ & 3,1,2,6,8,9    & SMP 32   & 0.00           & 2         \\
SMP 7  & $0.15\pm 0.09$ & 3,4,2,6        & SMP 34   & $0.10\pm 0.09$ & 2,9       \\
SMP 8  & $0.27\pm 0.25$ & 3,1,2,8,9      & MGPN 1   & 2.38           & 1         \\
SMP 9  & $0.93\pm 0.02$ & 3,1            & MGPN 2   & 2.87           & 1         \\
SMP 10 & $0.52\pm 0.42$ & 1,2,8          & MGPN 3   & 0.16           & 1         \\
SMP 11 & $0.56\pm 0.26$ & 3,1,7,2,6,8,9  & MGPN 5   & 0.00           & 1         \\
SMP 12 & $0.14\pm 0.19$ & 3,2,9          & MGPN 6   & 0.26           & 5         \\
SMP 13 & $0.30\pm 0.21$ & 3,1,2,6,8,9    & MGPN 7   & $0.99\pm 0.01$ & 1,2       \\
SMP 14 & $0.16\pm 0.11$ & 3,2,6,8,9      & MGPN 8   & 0.00           & 5         \\
SMP 15 & $0.16\pm 0.12$ & 3,1,2          & MGPN 9   & 1.08           & 2         \\
SMP 16 & $0.02\pm 0.02$ & 3,2            & MGPN 10  & 0.12           & 5         \\
SMP 17 & $0.65\pm 0.24$ & 3,2,8          & MGPN 11  & 1.72           & 1         \\
SMP 18 & $0.32\pm 0.16$ & 3,1,8,9        & MGPN 12  & 0.56           & 2         \\
SMP 19 & $0.37\pm 0.23$ & 3,2,8,9        & MGPN 13  & $0.34\pm 0.08$ & 1,2       \\
SMP 20 & $0.03\pm 0.03$ & 1,2,6          & Ma 01    & 0.00           & 5         \\
SMP 21 & $0.24\pm 0.18$ & 3,1,6          & Ma 02    & 0.99           & 5         \\
SMP 22 & $0.29\pm 0.14$ & 3,2,6,9        & [M95] 8  & 0.63           & 8         \\
SMP 23 & $0.05\pm 0.06$ & 3,2,6,9        & [M95] 9  & 0.24           & 3         \\
\hline
\end{tabular}
\begin{list}{}{}
\item[$^{*}$] References of individual coefficients
\item[] (1) This work\
\item[] (2) Leisy (\cite{leisy2}), Leisy \& Dennefeld (\cite{leisy3})\
\item[] (3) Costa et al. (\cite{costa})\
\item[] (4) Meyssonnier (\cite{meyssonier})\
\item[] (5) Vassiliadis et al. (\cite{vassiliadis})\
\item[] (6) Meatheringham \& Dopita (\cite{meatheringham1}, \cite{meatheringham2})\
\item[] (7) Boronson \& Liebert (\cite{boronson})\
\item[] (8) Monk et al. (\cite{monk})\
\item[] (9) Stanghellini et al. (\cite{stanghellini})\
\end{list}
\end{table*}
%References to the table

%
%_________________________________________________________________________________

\section{Physical parameters}

\begin{figure}
   \centering
   \includegraphics[angle=-90, width = 8cm]{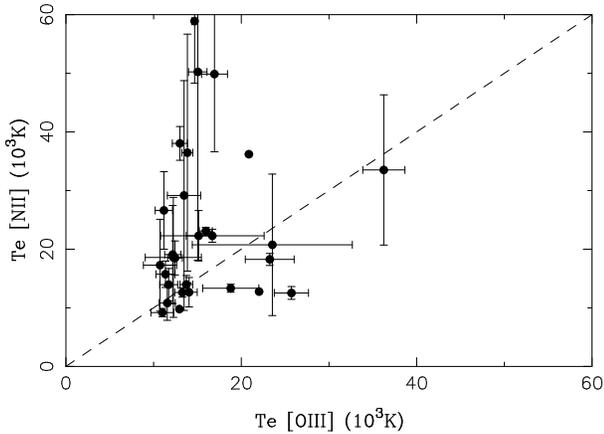}
   \caption{Electron temperatures  $T_e$ of PNe estimated  from [NII] 
       and [OIII] lines. Error bars are provided only for objects having
       more than one temperature determination.}
   \label{temperature}
   \end{figure}

Mean electron densities $N_e$ were estimated from the [SII] line ratio $\lambda$6716/6730. 
The observational dispersions in [SII] line fluxes were taken into account for $N_e$ 
calculations, giving a range of possible solutions for [SII] line ratios between 0.45 and 1.44. 
Mean electron temperatures $T_e$ were derived from the lines ratios [NII] 
($\lambda$6548+6583)/5755 and [OIII] ($\lambda$4959+5007)/4363. Table~\ref{parameters} shows 
the estimated mean electron densities (cm$^{-3}$) and temperatures (K) for each PNe.  
Errors were estimated by using classical error propagation both for electron densities and 
temperatures, leading to typical uncertainties of up to a factor of 2–-3 in densities and of 
10\% to 30\% in  the temperatures. It should be noted, however, that a proper account for 
the true accuracy of the derived physical parameters has to consider the number of flux 
measurements for each line. Those with few measurements may induce artificially low dispersions 
in their final values. In the case of a single measurement, again no error estimates are given.
When the [SII] line ratio was outside the range 0.45--1.44, upper and lower electron density 
limits of $N_e = 50000\,$cm$^{-3}$ and 10 cm$^{-3}$ respectively, were considered.
Fig.~\ref{temperature} shows the final estimated $T_e$ from [NII] and [OIII] lines.
Error bars are plotted only for objects having more than one temperature
determination. For those objects with a single measurement, an uncertainty of up to
30\% was estimated, as mentioned. For most nebulae there is a general agreement between 
the temperatures, in the sense that they do not differ by more than about 50\%, a result
similar to that obtained by Kingsburgh \& Barlow (\cite{kingsburgh}) for Galactic PNe.
However, the ratio of the [NII] to [OIII] temperatures depends on the excitation conditions,
and we find that for about 1/3 of the objects in the sample, the temperatures derived from 
[NII] lines are appreciably higher, which probably reflects the physical conditions in 
different parts of the nebulae. For the estimate of abundances, we decided to use $T_e$ 
from [OIII] lines due to the larger uncertainties in the [NII] fluxes.

%
%--------------------------------------------------------------------------------
%
% para fazer essa figura,(i) usei o acrobat 5 na figura 1 da
% Thais e salvei como jpg, (ii) transformei o jpg em ps com o 
% jpeg2ps, e (iii) fiz o eps com o gsview.
% 
%--------------------------------------------------------------------------------

%
% TABLE 5: physical parameters *** IN THE PAPER *** -------------------------------------------
%

\begin{table*}
\caption{Physical parameters: electron densities and temperatures.}
\label{parameters}
%      \centering
\begin{tabular}{lclllcll}
\hline\hline
Object    & $N_e$[SII]  &  $T_e$[NII]  & $T_e$[OIII] &\ Object  & $N_e$[SII]  &  $T_e$[NII]  & $T_e$[OIII] \\ 
\hline
SMP 1   & $>50000$  & $15720 \pm 2260$ & $11360\pm 1090$ &\ SMP 24   & 1240  & $13940\pm 2760$  & $11720\pm 1030$ \\
SMP 2   & 2740      & $12650 \pm 2500$ & $14040\pm 930$  &\ SMP 25   & 740   & $33510\pm 12820$ & $36260\pm 2400$ \\
SMP 3   & 4910      &                  & $12990\pm 610$  &\ SMP 26   & 840   & $13330\pm 680$   & $18800\pm 3200$ \\
SMP 4   & 1470      & $50230\pm 32120$ & $15040\pm 1050$ &\ SMP 27   & 11250 & $18590\pm 10230$ & $12250\pm 3200$ \\
SMP 5   & 4630      & $12600\pm 790$   & $13280\pm 860$  &\ SMP 28   & 2360  & $18250\pm 1020$  & $23250\pm 2790$ \\
SMP 6   & $>50000$  & $13960\pm 1550$  & $13740\pm 720$  &\ SMP 32   & 1790  &                  & 14760           \\
SMP 7   & 1790      & $49860\pm 13260$ & $16950\pm 1480$ &\ SMP 34   & 490   & 9770             & 12940           \\
SMP 8   & 1790      & $19070\pm 8380$  & $12180\pm 920$  &\ MGPN 1   & 460   & $58890\pm 10570$ & $14680\pm 330$  \\	   
SMP 9   & 790       & $36460\pm 20210$ & $13850\pm 620$  &\ MGPN 2   & 500   &                  & $25440\pm 2060$ \\
SMP 10  & 895       & $26610\pm 6620$  & $11180\pm 1000$ &\ MGPN 3   & 580   &                  & $17970\pm 330$  \\
SMP 11  & 1190      & $22290\pm 4300$  & $15140\pm 1410$ &\ MGPN 5   & 100   &                  & $17700\pm 6020$ \\
SMP 12  & 580       &                  & $13300\pm 760$  &\ MGPN 6   & 580   &                  & 85370           \\
SMP 13  & 1730      &                  & $12300\pm 1320$ &\ MGPN 7   & 445   & $23090\pm 650$   & $15960\pm 740$  \\
SMP 14  & 1470      &                  & $13450\pm 590$  &\ MGPN 8   & $>50000$ &               & 23900           \\
SMP 15  & 3350      & $18500\pm 2910$  & $12440\pm 840$  &\ MGPN 9   & 4010   & 36200           & 20870           \\
SMP 16  & $>50000$  & $9180\pm 700$    & $11000\pm 1300$ &\ MGPN 10  & 580    &                 & 25820           \\
SMP 17  & $>50000$  & $10810\pm 2950$  & $11560\pm 920$  &\ MGPN 11  & $<10$  & $22290\pm 1100$ & $16700\pm 5900$ \\
SMP 18  & 6020      & $17260\pm 7840$  & $10740\pm 1900$ &\ MGPN 12  & 1170   & 12760           & 22030           \\
SMP 19  & 4510      & $29150\pm 19590$ & $13470\pm 1900$ &\ MGPN 13  & 2470   &                 & $20130\pm 3620$ \\
SMP 20  & 1470      & $38040\pm 2870$  & $13000\pm 860$  &\ Ma 01    & 580    &                 & 19690           \\	   
SMP 21  & 11250     & $20740\pm 12060$ & $23540\pm 9120$ &\ Ma 02    & 580    &                 & 20990           \\
SMP 22  & 1660      & $12530\pm 1080$  & $25720\pm 1950$ &\ [M95] 8  &        &                 &                 \\
SMP 23  & 580       &                  & $13265\pm 319$  &\ [M95] 9  & 230    &                 & $12400\pm 740$  \\
\hline
\end{tabular}
\end{table*}

%
%_________________________________________________________________________________

\section{Abundances}

\subsection{Helium abundances}

Helium abundances were estimated using the mean electron temperatures and densities derived in 
the previous section. In view of their relatively intense flux, the lines used to derive the 
He abundance were HeI $\lambda$5876 and HeII $\lambda$4686. The total helium abundance is then 
given by 

   \begin{equation}
   {{\rm He} \over {\rm H}} = {\alpha_{5876} \ X^2_{5876} + \alpha_{4676} \ X^2_{4676} \over
        \alpha_{4861} \ X^2_{4861}} \,
   \end{equation}

\noindent  
where $\alpha_\lambda$ are the total recombination coefficients and $X_\lambda$ represent the 
fluxes in units of the H$\beta$ line flux. The total recombination coefficients are taken from 
P\'equignot et al. (\cite{daniel}). In the case of HeI lines, corrections of important 
collisional effects were made in the recombination coefficients, using the 
collision-to-recombination correction factors from Kingdon \& Ferland (\cite{kingdon}).
 
Table~\ref{helium} shows the derived ionic and total helium abundances relative to
hydrogen, by number of atoms. In this table and in the next one the notation 
$2 \times 10^{-4} = 2(-4)$ has been used. Errors were estimated by propagating the observational 
uncertainties in HeI and HeII line fluxes and in the $T_e$ and $N_e$ derivation. In the case of MG7 
the He abundance was calculated from the $\lambda$4471 line. For MG8 and MG13 average values 
have been used.

%
% TABLE 6: Helium abundances  *** IN THE PAPER *** -------------------------------------------
%
\begin{table*}
\caption{Helium ionic and total abundances.}
\label{helium}
%      \centering
\begin{tabular}{lclllcll}
\hline\hline
Object    & HeI           &  HeII             & He                &\ Object&  HeI             &  HeII              & He \\
\hline
SMP 1  & $0.071\pm 0.012$  & $1(-3)\pm 2(-4)$  & $0.072\pm 0.012$  &\ SMP 24  & $0.097\pm 0.011$ &                  & $0.097\pm 0.011$ \\
SMP 2  & $0.072\pm 0.008$  & $0.018\pm 0.003$  & $0.090\pm 0.009$  &\ SMP 25  & $0.057\pm 0.006$ & $4(-3)\pm 9(-4)$ & $0.061\pm 0.006$ \\
SMP 3  & $0.114\pm 0.001$  & $7(-3)\pm 8(-4)$  & $0.121\pm 0.001$  &\ SMP 26  & $0.058\pm 0.017$ & $0.018\pm 0.002$ & $0.075\pm 0.017$ \\
SMP 4  & $0.124\pm 0.011$  & $2(-4)\pm 2(-4)$  & $0.125\pm 0.011$  &\ SMP 27  & $0.091\pm 0.009$ & $1(-3)\pm 3(-5)$ & $0.092\pm 0.009$ \\ 
SMP 5  & $0.069\pm 0.009$  & $0.026\pm 0.002$  & $0.095\pm 0.009$  &\ SMP 28  & $0.089\pm 0.009$ & $12(-3)\pm 6(-4)$& $0.100\pm 0.009$ \\ 
SMP 6  & $0.087\pm 0.010$  & $3(-4)\pm 4(-6)$  & $0.088\pm 0.010$  &\ SMP 32  & $0.085\pm 0.007$ & 0.006            & $0.091\pm 0.007$ \\
SMP 7  & $0.080\pm 0.008$  & $4(-3)\pm 7(-4)$  & $0.084\pm 0.008$  &\ SMP 34  & $0.079\pm 0.008$ & 0.006            & $0.085\pm 0.008$ \\
SMP 8  & $0.099\pm 0.014$  &                   & $0.099\pm 0.014$  &\ MGPN 1  & $0.151\pm 0.023$ & $11(-3)\pm 3(-4)$& $0.161\pm 0.023$ \\
SMP 9  & $0.030\pm 0.004$  & $0.034\pm 0.006$  & $0.064\pm 0.007$  &\ MGPN 2  & $0.067\pm 0.026$ & $2(-3)\pm 1(-4)$ & $0.069\pm 0.026$ \\
SMP 10 & $0.087\pm 0.011$  & $6(-4)\pm 6(-4)$  & $0.088\pm 0.011$  &\ MGPN 3  & $0.129\pm 0.014$ & $7(-3)\pm 2(-4)$ & $0.136\pm 0.014$ \\
SMP 11 & $0.085\pm 0.010$  & $1(-3)\pm 8(-5)$  & $0.086\pm 0.010$  &\ MGPN 5  & $0.067\pm 0.038$ & $0.034\pm 0.002$ & $0.101\pm 0.038$ \\
SMP 12 & $0.107\pm 0.009$  & $1(-3)\pm 3(-4)$  & $0.108\pm 0.009$  &\ MGPN 6  & $0.149\pm 0.012$ & 0.002            & $0.151\pm 0.012$ \\
SMP 13 & $0.107\pm 0.015$  & $2(-4)\pm 7(-6)$  & $0.107\pm 0.015$  &\ MGPN 7  & $0.104\pm 0.050$ & $0.013\pm 0.005$ & $0.117\pm 0.050$ \\
SMP 14 & $0.062\pm 0.007$  & $0.021\pm 0.004$  & $0.084\pm 0.008$  &\ MGPN 8  & $0.019\pm 0.006$ & $5(-3)\pm 6(-5)$ & $0.024\pm 0.006$ \\
SMP 15 & $0.104\pm 0.011$  & $5(-4)\pm 3(-4)$  & $0.105\pm 0.011$  &\ MGPN 9  & $0.063\pm 0.007$ & $1(-4)$          & $0.063\pm 0.007$ \\
SMP 16 & $0.062\pm 0.008$  & $6(-4)\pm 3(-4)$  & $0.063\pm 0.008$  &\ MGPN 10 & $0.070\pm 0.006$ & 0.019            & $0.089\pm 0.006$ \\
SMP 17 & $0.103\pm 0.024$  & $1(-3)\pm 2(-4)$  & $0.104\pm 0.024$  &\ MGPN 11 & $0.218\pm 0.036$ & $0.008\pm 0.004$ & $0.227\pm 0.036$ \\
SMP 18 & $0.086\pm 0.008$  &                   & $0.086\pm 0.008$  &\ MGPN 12 & $0.118\pm 0.010$ & 0.013            & $0.131\pm 0.010$ \\
SMP 19 & $0.055\pm 0.008$  & $0.026\pm 0.003$  & $0.081\pm 0.009$  &\ MGPN 13 & $0.006\pm 0.002$ & $0.031\pm 0.003$ & $0.037\pm 0.003$ \\
SMP 20 & $0.118\pm 0.020$  & $2(-4)\pm 4(-6)$  & $0.119\pm 0.020$  &\ Ma 01   & $0.167\pm 0.014$ & 0.007            & $0.174\pm 0.014$ \\
SMP 21 & $0.075\pm 0.017$  & $0.007\pm 0.001$  & $0.082\pm 0.017$  &\ Ma 02   & $0.049\pm 0.004$ & 0.011            & $0.061\pm 0.004$ \\
SMP 22 & $0.062\pm 0.007$  & $11(-3)\pm 9(-4)$ & $0.073\pm 0.007$  &\ [M95] 8 &                  &                  &                  \\
SMP 23 & $0.095\pm 0.011$  & $2(-3)\pm 2(-4)$  & $0.097\pm 0.011$  &\ [M95] 9 & $0.083\pm 0.007$ &                  & $0.083\pm 0.007$ \\
\hline
\end{tabular}
\end{table*}

\subsection{Ionic and elemental abundances}

Ionic abundances were estimated for the ions present in the optical spectra by solving the 
statistical equilibrium equations for a three-level atom model, including radiative and 
collisional transitions. The resulting ionic abundances with uncertainties are given in 
Table~7 (available at the CDS). In this table, the first line of each nebula gives the ionic abundances by 
number of atoms relative to hydrogen, while the second line gives the uncertainties. The 
asterisks (*) indicate that the corresponding dispersion is due to different measurements 
in electronic temperature and/or densities, while double asterisks (**) indicate that the 
dispersions are due to different measurements in the ionic line only. The sign (:) indicates 
only one measurement of the electron temperature, density and ionic line, so that no errors 
are given. The elemental abundances were then derived using ionization correction factors 
(ICF).  We adopted the same ICF given by Escudero et al. (\cite{escudero}) to account for 
unobserved ions of each element. The corresponding elemental abundances are given in columns~2 
to 7 of Table~\ref{elemental}, in the form $\epsilon({\rm X}) = \log ({\rm X/H}) + 12$, as usual.  
It should be noted that the uncertainties given in Table~\ref{elemental} are formal
uncertainties, which are essentially an estimate of the dispersion in the considered measurements,
as discussed in Section~2. A more realistic estimate of the errors may be obtained from a 
comparison with other determinations in the literature (cf. section 4.4), and may reach about 
0.1 to 0.2 dex for the best measured elements, and about 0.3 dex for those with the weakest lines.

In order to derive the carbon abundances we adopted the following procedure: ionic abundances 
of carbon and oxygen were derived from the ultraviolet CIII]$\lambda$1909 and OIII]$\lambda$1663 
lines, according to the formulae by Aller (\cite{aller2}). Rola \& Stasi\'nska (\cite{rola}) 
pointed out that the elemental ratio C/O can be safely approximated by the C$^{+2}$/O$^{+2}$ 
ratio, except for very low excitation objects (those with [OIII]$\lambda$5007/H$\beta < 3$),
for which this assumption is not valid. We derived the C/O ratio, and, by taking the oxygen 
abundance derived from the optical data, the carbon abundances were 
derived for the nebulae. However, one should keep in mind that, as concluded by Rola \& 
Stasi\'nska (\cite{rola}), the C/O ratio for a given object may present discrepancies according 
to the lines (in the optical and/or UV range) used. In particular, when derived only from UV lines, 
it tends to be underestimated when compared to values derived from optical and UV lines. We 
adopted the procedure described above, mainly due to the uncertainties implicit in the reddening 
determination that can lead to large errors when UV and optical lines are combined.

%
% TABLE 7:  Ionic abundances *** CDS *** ----------------------------------------------------------
%

% TABLE 8: elemental abundances *** IN THE PAPER ***  --------------------------------------------
\setcounter{table}{7}
\begin{table*}
\caption{Elemental abundances.}  
\label{elemental} 
%      \centering
\begin{tabular}{lccccccc}
\hline\hline
Object    & C &  N & S & Ar & Ne & O & $Z$  \\
\hline
SMP 1  & $8.11\pm 0.26$ & $7.08\pm 0.12$ & $6.37\pm 0.03$ & $6.04\pm 0.11$ & $7.57\pm 0.04$ & $8.26\pm 0.11$ & $0.006\pm 0.001$ \\
SMP 2  & $7.99\pm 0.25$ & $7.59\pm 0.07$ & $7.09\pm 0.02$ & $5.62\pm 0.07$ & $7.30\pm 0.02$ & $8.20\pm 0.07$ & $0.005\pm 0.001$ \\
SMP 3  & $7.98\pm 0.25$ & $7.28\pm 0.08$ & $7.13\pm 0.02$ & $5.40\pm 0.08$ & $7.36\pm 0.01$ & $8.20\pm 0.08$ & $0.005\pm 0.001$ \\
SMP 4  & $7.45\pm 0.25$ & $7.48\pm 0.07$ & $7.68\pm 0.02$ & $5.25\pm 0.07$ & $7.15\pm 0.01$ & $7.91\pm 0.07$ & $0.004\pm 0.001$ \\
SMP 5  & $8.36\pm 0.25$ & $7.72\pm 0.07$ & $7.50\pm 0.02$ & $5.89\pm 0.07$ & $7.48\pm 0.02$ & $8.39\pm 0.07$ & $0.009\pm 0.002$ \\
SMP 6  & $8.03\pm 0.25$ & $7.66\pm 0.08$ & $6.80\pm 0.03$ & $5.81\pm 0.08$ & $7.46\pm 0.02$ & $8.22\pm 0.07$ & $0.005\pm 0.001$ \\
SMP 7  & $7.54\pm 0.25$ & $7.31\pm 0.08$ & $7.00\pm 0.01$ & $5.30\pm 0.09$ & $7.26\pm 0.02$ & $7.96\pm 0.07$ & $0.003\pm 0.001$ \\
SMP 8  & $7.78\pm 0.26$ & $7.04\pm 0.10$ & $7.11\pm 0.02$ & $5.48\pm 0.09$ & $7.23\pm 0.02$ & $8.09\pm 0.09$ & $0.004\pm 0.001$ \\
SMP 9  & $8.58\pm 0.26$ & $7.69\pm 0.09$ & $6.94\pm 0.03$ & $6.07\pm 0.08$ & $7.51\pm 0.02$ & $8.51\pm 0.08$ & $0.011\pm 0.003$ \\
SMP 10 & $8.18\pm 0.25$ & $7.80\pm 0.10$ & $7.33\pm 0.05$ & $5.66\pm 0.09$ & $7.50\pm 0.03$ & $8.30\pm 0.08$ & $0.007\pm 0.001$ \\
SMP 11 & $7.11\pm 0.25$ & $6.52\pm 0.08$ & $6.42\pm 0.02$ & $5.86\pm 0.08$ & $6.91\pm 0.03$ & $7.74\pm 0.08$ & $0.001\pm 0.001$ \\
SMP 12 & $7.48\pm 0.25$ & $7.29\pm 0.07$ & $8.16\pm 0.04$ & $5.05\pm 0.07$ & $7.11\pm 0.01$ & $7.93\pm 0.06$ & $0.007\pm 0.001$ \\
SMP 13 & $8.03\pm 0.26$ & $7.79\pm 0.10$ & $7.41\pm 0.03$ & $5.53\pm 0.09$ & $7.36\pm 0.02$ & $8.22\pm 0.09$ & $0.006\pm 0.001$ \\
SMP 14 & $8.15\pm 0.25$ & $7.65\pm 0.07$ & $7.11\pm 0.02$ & $5.64\pm 0.07$ & $7.44\pm 0.01$ & $8.29\pm 0.07$ & $0.006\pm 0.001$ \\
SMP 15 & $7.73\pm 0.25$ & $7.80\pm 0.08$ & $7.37\pm 0.03$ & $5.58\pm 0.07$ & $7.41\pm 0.01$ & $8.06\pm 0.07$ & $0.005\pm 0.001$ \\
SMP 16 & $8.89\pm 0.26$ & $7.11\pm 0.10$ & $6.55\pm 0.02$ & $6.31\pm 0.10$ & $8.03\pm 0.05$ & $8.67\pm 0.09$ & $0.019\pm 0.006$ \\
SMP 17 & $8.63\pm 0.28$ & $7.06\pm 0.15$ & $6.68\pm 0.02$ & $5.84\pm 0.15$ & $7.73\pm 0.03$ & $8.54\pm 0.14$ & $0.012\pm 0.004$ \\
SMP 18 & $7.68\pm 0.25$ & $7.32\pm 0.15$ & $6.54\pm 0.03$ & $5.82\pm 0.08$ & $7.46\pm 0.05$ & $8.04\pm 0.07$ & $0.003\pm 0.001$ \\
SMP 19 & $8.16\pm 0.26$ & $7.79\pm 0.11$ & $7.16\pm 0.03$ & $5.40\pm 0.09$ & $7.34\pm 0.03$ & $8.29\pm 0.09$ & $0.007\pm 0.001$ \\
SMP 20 & $7.30\pm 0.26$ & $7.44\pm 0.11$ & $7.21\pm 0.02$ & $5.01\pm 0.11$ & $7.06\pm 0.01$ & $7.84\pm 0.11$ & $0.002\pm 0.001$ \\
SMP 21 & $6.59\pm 0.28$ & $7.85\pm 0.20$ & $6.71\pm 0.05$ & $5.42\pm 0.15$ & $6.88\pm 0.05$ & $7.46\pm 0.14$ & $0.002\pm 0.001$ \\
SMP 22 & $6.06\pm 0.25$ & $7.96\pm 0.06$ & $6.30\pm 0.01$ & $5.23\pm 0.06$ & $6.61\pm 0.02$ & $7.18\pm 0.06$ & $0.002\pm 0.001$ \\
SMP 23 & $7.80\pm 0.25$ & $7.12\pm 0.08$ & $7.39\pm 0.01$ & $5.45\pm 0.07$ & $7.39\pm 0.01$ & $8.10\pm 0.07$ & $0.004\pm 0.001$ \\
SMP 24 & $7.69\pm 0.25$ & $7.26\pm 0.08$ & $6.59\pm 0.02$ & $5.63\pm 0.08$ & $7.34\pm 0.02$ & $8.04\pm 0.07$ & $0.003\pm 0.001$ \\
SMP 25 & $5.63\pm 0.26$ & $7.72\pm 0.10$ & $6.72\pm 0.02$ & $4.92\pm 0.09$ & $6.43\pm 0.02$ & $6.96\pm 0.09$ & $0.001\pm 0.001$ \\
SMP 26 & $7.27\pm 0.29$ & $7.85\pm 0.17$ & $6.28\pm 0.03$ & $5.57\pm 0.16$ & $7.22\pm 0.04$ & $7.82\pm 0.16$ & $0.003\pm 0.001$ \\
SMP 27 & $7.77\pm 0.25$ & $7.09\pm 0.17$ & $6.72\pm 0.05$ & $5.33\pm 0.08$ & $7.26\pm 0.05$ & $8.08\pm 0.07$ & $0.003\pm 0.001$ \\
SMP 28 & $6.10\pm 0.25$ & $8.04\pm 0.07$ & $6.67\pm 0.03$ & $5.38\pm 0.07$ & $6.78\pm 0.02$ & $7.21\pm 0.06$ & $0.002\pm 0.001$ \\
SMP 32 & $7.53\pm 0.24$ & $7.09\pm 0.05$ & $7.98$         & $4.78\pm 0.05$   & $7.15$       & $7.96\pm 0.05$ & $0.005\pm 0.001$ \\
SMP 34 & $7.71\pm 0.25$ & $7.24\pm 0.06$ & $6.06$         & $5.48\pm 0.06$   & $7.53$       & $8.05\pm 0.06$ & $0.003\pm 0.001$ \\
MGPN 1   & $7.51\pm 0.26$ & $7.95\pm 0.09$ & $8.01\pm 0.01$ & $5.85\pm 0.10$ & $6.68$       & $7.95\pm 0.09$ & $0.006\pm 0.001$ \\
MGPN 2   & $6.53\pm 0.34$ & $6.33\pm 0.24$ & $6.90\pm 0.01$ & $5.49\pm 0.24$ & $6.88\pm 0.04$& $7.43\pm 0.24$& $0.001\pm 0.001$ \\
MGPN 3   & $7.01\pm 0.26$ & $6.02\pm 0.24$ & $7.37\pm 0.11$ & $5.80\pm 0.14$ & $6.98\pm 0.12$& $7.68\pm 0.10$& $0.002\pm 0.001$ \\
MGPN 5   & $6.95\pm 0.39$ & $7.52\pm 0.33$ & $7.74\pm 0.20$ & $5.98\pm 0.31$ & $6.69\pm 0.11$& $7.65\pm 0.30$& $0.003\pm 0.001$ \\
MGPN 6   & $5.12\pm 0.25$ & $6.74\pm 0.05$ & $7.07$         & $5.51\pm 0.05$ & $5.81$        & $6.69\pm 0.05$& $0.001\pm 0.001$ \\
MGPN 7   & $7.45\pm 0.44$ & $7.28\pm 0.36$ & $5.90\pm 0.02$ & $5.61\pm 0.36$ & $7.63\pm 0.01$& $7.92\pm 0.36$& $0.003\pm 0.001$ \\
MGPN 8   & $8.23\pm 0.30$ & $7.02\pm 0.18$ & $6.06$         & $6.74\pm 0.18$ & $7.59$        & $8.33\pm 0.18$& $0.007\pm 0.002$ \\
MGPN 9   & $6.47\pm 0.25$ & $6.64\pm 0.07$ & $6.57$         & $5.28\pm 0.07$ & $6.70$        & $7.40\pm 0.07$& $0.001\pm 0.001$ \\
MGPN 10  & $5.97\pm 0.24$ & $7.07\pm 0.05$ & $7.51$         & $5.53\pm 0.05$ & $6.34$        & $7.14\pm 0.05$& $0.001\pm 0.001$ \\
MGPN 11  & $7.47\pm 0.29$ & $7.85\pm 0.34$ & $7.43\pm 0.21$ & $5.76\pm 0.20$ & $6.67\pm 0.14$& $7.93\pm 0.17$& $0.004\pm 0.001$ \\
MGPN 12  & $6.23\pm 0.24$ & $8.22\pm 0.05$ & $6.50$         & $5.57\pm 0.05$ & $6.97$        & $7.27\pm 0.05$& $0.003\pm 0.001$ \\
MGPN 13  & $7.66\pm 0.27$ & $6.54\pm 0.16$ & $6.18\pm 0.03$ & $5.63\pm 0.13$ & $6.65\pm 0.03$& $8.03\pm 0.12$& $0.002\pm 0.001$ \\
Ma 01  & $7.07\pm 0.24$ & $7.64\pm 0.05$ & $8.27$         & $5.95\pm 0.05$ & $6.98$          & $7.72\pm 0.05$& $0.008\pm 0.001$ \\
Ma 02  & $7.04\pm 0.24$ & $6.87\pm 0.05$ & $6.43$         & $5.44\pm 0.05$ & $7.25$          & $7.70\pm 0.05$& $0.001\pm 0.001$ \\
{[M95] 8}    &                &                &                &                &                &                &                  \\
{[M95] 9}    &                &                &                &                &                &                &                  \\
\hline
\end{tabular}
\end{table*}

\subsection{The total heavy element abundance $Z$}

In order to extend our study to the PNe central stars, an estimate of the total heavy element
abundance by mass $Z$ is needed. We have initially obtained an estimate of $Z$ for the PNe 
in our sample adopting the procedure outlined by Chiappini \& Maciel (\cite{chiappini}), so 
that we may write

   \begin{equation}
    Z \simeq {\sum A_i \, (n_i/n_H) \over 1 + 4 \, ({\rm He/H}) + \sum A_i \, (n_i/n_H)}  \,
   \end{equation}

\noindent
where $A_i$ and $n_i/n_H$ are the mass number and abundance of element $i$ relative
to H, respectively, and the sum includes the metals ($A_i > 2$) for which detailed 
abundances are available. Some of the observed elements (He, C, N) are contaminated
during the evolution of the PNe central star, which introduces an error in the estimate
of $Z$. In order to overcome such difficulty, we have considered an alternative approach
based on a correlation between the heavy element and the oxygen abundances (cf. Chiappini 
\& Maciel \cite{chiappini}). Additionally,  we have attempted to correct the heavy 
element abundances given by Eq.~2 by assuming that a fraction of the observed abundances
of these elements has in fact been produced by nucleosynthetic processes in the 
progenitor stars. It turns out that the different estimates of the heavy element
abundances are very similar, with average deviations $\Delta Z \simeq 0.001$ to 0.002,
which is probably due to the dominant role of oxygen, which is not enhanced in the
progenitor stars. The adopted heavy element abundances $Z$ are given in the
last column of Table~\ref{elemental}.  

\subsection{Comparison with previous results}

Several works have dealt with the determination of the chemical composition
of PNe in the SMC, so it is interesting to compare our results with some
previous abundance determinations. Stasi\'nska et al. (\cite{stasinska})
have collected photometric and spectroscopic data on PNe in five galaxies
including the SMC, a work later extended by Richer \& McCall (\cite{richer}).
In the SMC, the PNe sample is similar to our present sample, containing about
60 objects. From Stasi\'nska et al. (\cite{stasinska}), an average oxygen
abundance $\langle{\epsilon_O}\rangle = \langle \log{\rm O/H}+12 \rangle 
\simeq 7.74$ to $8.10$ was obtained, depending on the luminosity of the objects, 
in excellent agreement with our own average, $\langle{\epsilon_O}\rangle \simeq
7.89$, estimated from Table~\ref{elemental}. Also for the N/O ratio, Stasi\'nska
et al. (\cite{stasinska}) obtain an average in the range 
$-0.46 > \langle{\log {\rm N/O}}\rangle > -0.66$, while our average is 
$\langle{\log{\rm N/O}}\rangle \simeq -0.55$.

\begin{figure}
   \centering
   \includegraphics[angle=-90, width = 8cm]{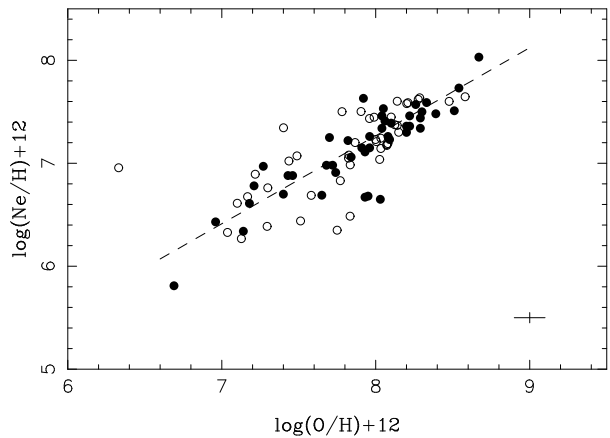}
   \caption{Neon and Oxygen abundances for PNe in the SMC. Filled circles: this
     work. Empty circles: data from Stasi\'nska et al. (\cite{stasinska}).
     The dashed line shows a least squares fit to our data. An average error
     bar is shown at the lower right corner.}
   \label{neon}
   \end{figure}

Planetary nebulae can be useful as probes of the chemical evolution of galaxies by
considering distance-independent correlations involving the measured abundances,
which can then be compared with predictions of chemical evolution models. Also,
these correlations are an efficient way to compare abundance determinations
from different sources. A recent discussion of some of these correlations for PNe 
in the Galaxy and in the Magellanic Clouds has been given by Maciel et al. 
(\cite{mci2006}), to which the reader is referred for details. Here we will 
present an example of such a correlation, for the elements Ne and O. Since these 
elements are not primarily produced in the PNe central stars, a tight correlation 
can be expected, as shown in Fig.~\ref{neon}. In this figure, filled circles show 
the present abundances, and the dashed line shows the corresponding least squares 
fit, with a slope of 0.86 and a correlation coefficient $r \simeq 0.92$.  
An average error bar according to the data in Table~\ref{elemental} is included 
at the lower right corner. 

%
%--------------------------------------------------------------------------------

% 
%--------------------------------------------------------------------------------

The results by Stasi\'nska et al. (\cite{stasinska}) are also included in 
Fig.~\ref{neon} as empty circles, and it can be concluded that there is a very
good agreement between the different sources. In fact, as shown by Richer \&
McCall (\cite{richer}), the Ne/H $\times$ O/H relation observed in bright PNe
is essentially the same found in the interstellar medium of star forming
galaxies, including several objects in the Local Group. The usual interpretation
of this fact is that the progenitors of most PNe do not significantly modify
either of these abundances. Other alpha-elements in our sample also show good 
correlations with oxygen, except for sulphur. This can be explained by the very 
low flux of the [SII] lines, which are good for the determination of the electron 
density, but not for ionic abundances. On the other hand, the sulphur anomaly as 
discussed by Henry et al. (\cite{henry}) may also play a part.

In a recent paper, Leisy \& Dennefeld (\cite{leisy3}) presented a detailed analysis
of a large sample of PNe in the Magellanic Clouds. The sample included their own
data plus other objects from the literature, for which the elemental abundances
were derived in a homogeneous way. It is interesting to compare  our results
with this sample, but it should be noted that the SMC sample of Leisy \& Dennefeld
(\cite{leisy3}) is actually smaller than ours, amounting to 37 objects with
derived abundances. They make an attempt to adapt to the SMC the Peimbert criteria 
(cf. Peimbert \cite{peimbert}),  which classify Galactic PNe according to
their chemical, spatial and kinematical properties. Such an adaptation is not
obvious, as it is generally based on a comparison of chemical abundances,
while the original Peimbert criteria also take into account the space and kinematical
properties, which are different in spiral and irregular galaxies. Furthermore, the
Magellanic Clouds have lower metallicity compared to the Galaxy, and some sort
of calibration needs to be done for PNe in the LMC and SMC, which introduces an
additional uncertainty. In fact, Leisy \& Dennefeld (\cite{leisy3}) do not obtain
a clearcut separation between their Type I and non-Type I objects, and find
a continuity in the He and N/O abundances between these types. Therefore, this
distinction is not well defined, so that in the following we will not give further 
consideration to this separation. Table~\ref{comparison}
shows our average results obtained from Table~\ref{elemental} in comparison with
the averages from Leisy \& Dennefeld (\cite{leisy3}) for the SMC objects. We
reproduce their results for Type I and non-Type I objects, as well as their 
compilation for HII region abundances, taken from Dennefeld (\cite{dennefeld}).

%
% TABLE 9: Abundance comparations  *** IN THE PAPER *** -------------------------------------------
%
\begin{table*}
\caption{Average abundances compared with Leisy \& Dennefeld (\cite{leisy3}).}
\label{comparison}
%      \centering
\begin{tabular}{lcccccccc}
\hline\hline
    & He  & C & N & O & Ne & S & Ar & N/O  \\
\hline
HII regions       & 10.90 & 7.19 & 6.46 & 7.97 & 7.22 & 6.32 & 5.78 & $-$1.51 \\
Type I (LD)       & 11.09 & 7.82 & 7.78 & 7.80 & 7.09 & 7.01 & 5.67 & $-$0.03 \\
non-Type I (LD)   & 10.90 & 8.81 & 7.11 & 8.09 & 7.16 & 6.62 & 5.51 & $-$0.98 \\
This work         & 10.99 & 7.41 & 7.35 & 7.89 & 7.14 & 6.98 & 5.59 & $-$0.55 \\
\hline
\end{tabular}
\end{table*}

Table~\ref{comparison} shows a very good agreement between our average 
abundances and those by Leisy \& Dennefeld (\cite{leisy3}), taken as 
an average between their Type I and non-Type I objects. Moreover, from a 
comparison of the PNe averages with those from HII regions, we do not find
any clear evidence of contamination in the progenitor stars of the elements
O, Ne, Ar, and S, suggesting that ON cycling may operate in the massive
progenitors only, which probably consist of a small fraction of the stars
leading to the PNe in our sample. Naturally, for He, C, and N there are important 
differences between the nebular gas in PNe and the interstellar gas, which are 
admittedly due to the nucleosynthetic processes in the progenitor stars.

\section{Effective temperatures and luminosities of the central stars}

To derive the effective temperature ($T_{eff}$) of PNe central stars, we used the method 
described by Kaler \& Jacoby (\cite{kaler}).  The temperatures of PNe progenitors were 
estimated using HeII $\lambda$4686/H$\beta$ ratio, which is a better $T_{eff}$ estimator 
than the [OIII] $\lambda$5007/H$\beta$ ratio, as discussed largely in the literature.

Progenitor luminosities $L/L_\odot$ were derived from Kaler \& Jacoby (\cite{kaler}) 
relations of $V$ magnitude with H$\beta$ absolute flux, $T_{eff}$ and extinction coefficient 
$\langle c_{ext} \rangle$. H$\beta$ absolute fluxes were taken from the Meatheringham et al. 
(\cite{meatheringham3}) compilation, complemented by Wood et al. (\cite{wood}) data as necessary. 
$\langle c_{ext} \rangle$ are from the compilation of Table~\ref{extinction}. The estimated errors 
for H$\beta$ fluxes and $\langle c_{ext} \rangle$ are basically the dispersion of the different 
measurements. When the measurements are from one source, the estimated  errors are given by the 
original references.  

\begin{figure*}
   \centering
   \includegraphics[angle=0, width = 13cm]{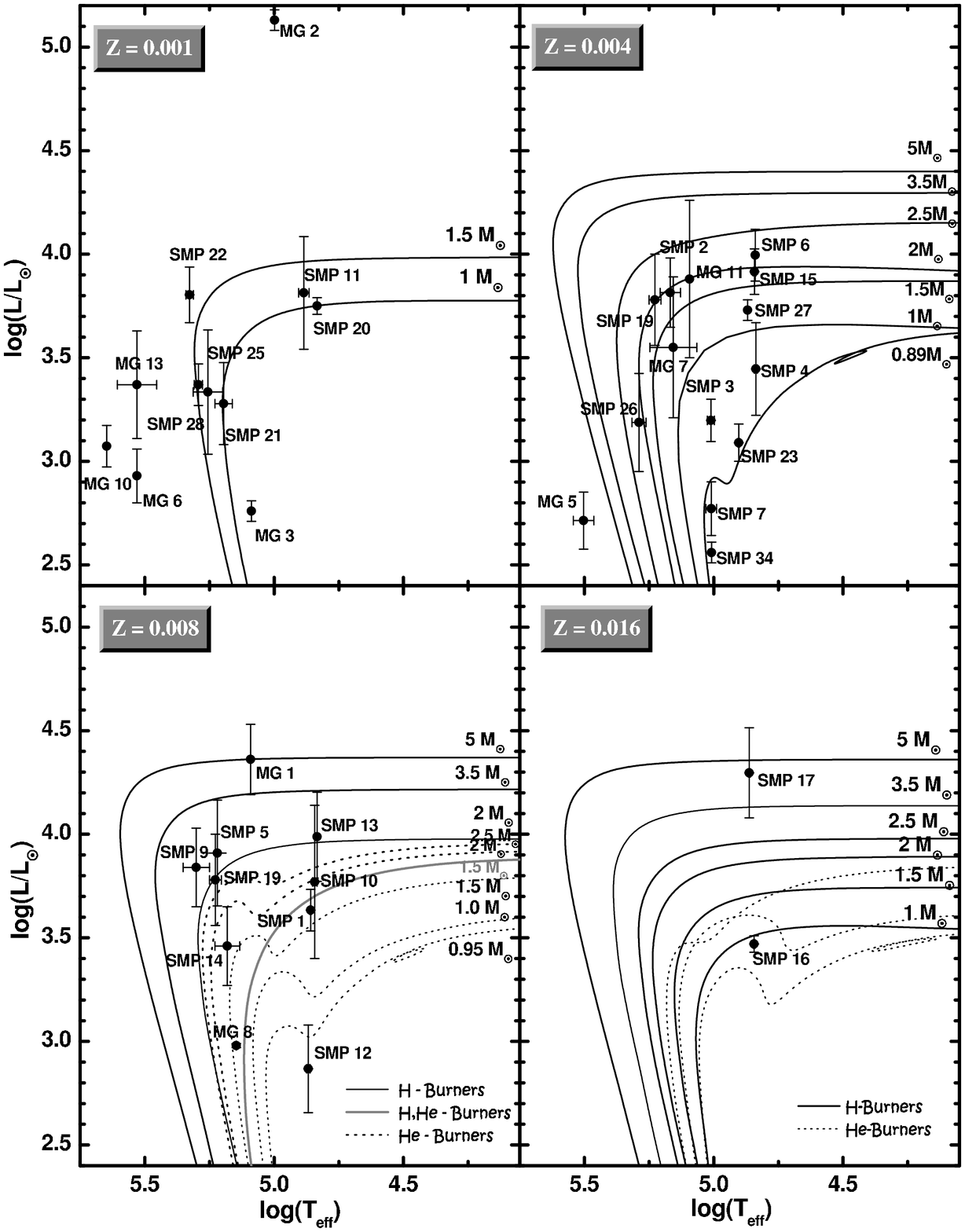}
   \caption{Position of the PNe central stars on the HR diagram.
    Isochrones are from Vassiliadis \& Wood (\cite{vassiliadis2}).}
   \label{hrdiagram}
   \end{figure*}

To transform $V$ into $L/L_\odot$, we used the relations involving the bolometric corrections 
as given by Cazetta \& Maciel (\cite{cazetta}), assuming a SMC distance of 57.5 kpc (Feast \& 
Walker \cite{feast}), which is within 5\% of most recent determinations (see for example
Harries et al. \cite{harries} and Keller \& Wood \cite{keller}).  Masses, and then ages, 
were derived using the Vassiliadis \& Wood (\cite{vassiliadis2}) isochrones and mass-age 
relationships, with the luminosities and effective temperatures derived as explained above. 
The isochrones were selected according to the derived heavy element abundance $Z$, as given 
in Table~\ref{elemental}. The resulting progenitor star masses and ages are given in 
Table~\ref{progenitor}, along with the effective temperatures and luminosities. 
Figure~\ref{hrdiagram} displays the position of each object on the HR diagram over the adopted 
isochrones. We found that most central stars are younger than about 6 Gyr, which is similar to 
the results for galactic PNe from Maciel et al. (\cite{mqc2007}), but in the case of the SMC 
the fraction of objects younger than 4 Gyr is much higher than in the case of the Galaxy.

% TABLE 10: progenitor parameters  *** IN THE PAPER ***   -----------------------------------------
\begin{table*}
\caption{Progenitor star parameters$^a$.} 
\label{progenitor}
%      \centering
\begin{tabular}{lcccc}
\hline\hline
Object    & $\log T_{eff}$ & $\log (L/L_\odot)$ & Mass ($M_\odot$) & Age (Gyr)  \\
\hline
SMP 1   & $4.859\pm 0.003$ & $3.63\pm 0.10$ & $1.7\pm 0.2$   & 1.85 \\
SMP 2   & $5.168\pm 0.039$ & $3.81\pm 0.17$ & $1.9\pm 0.5$   & 1.39 \\
SMP 3   & $5.011\pm 0.014$ & $3.20\pm 0.10$ & $0.95\pm 0.02$ & 12.2 \\
SMP 4   & $4.836\pm 0.006$ & $3.45\pm 0.22$ & $0.98\pm 0.11$ & 10.6 \\
SMP 5   & $5.219\pm 0.017$ & $3.91\pm 0.26$ & $2.3\pm 0.3$   & 0.83 \\
SMP 6   & $4.839$          & $4.00\pm 0.12$ & $2.1\pm 0.5$   & 1.03 \\
SMP 7   & $5.009\pm 0.020$ & $2.77\pm 0.13$ & $0.89$         &      \\
SMP 8   &                  &                &                &      \\
SMP 9   & $5.301\pm 0.051$ & $3.84\pm 0.19$ & $2.5\pm 0.5$   & 0.69 \\
SMP 10  & $4.843\pm 0.009$ & $3.77\pm 0.37$ & $2.0\pm 0.7$   & 1.05 \\
SMP 11  & $4.885\pm 0.020$ & $3.81\pm 0.27$ & $1.2$          & 7.0  \\
SMP 12  & $4.868\pm 0.006$ & $2.87\pm 0.21$ & $0.95^*$       & 11.4 \\
SMP 13  & $4.834$          & $3.99\pm 0.22$ & $2.5\pm 0.7$   & 0.69 \\
SMP 14  & $5.182\pm 0.048$ & $3.46\pm 0.19$ & $1.7\pm 0.2$   & 1.85 \\
SMP 15  & $4.843\pm 0.005$ & $3.92\pm 0.11$ & $1.9\pm 0.5$   & 1.39 \\
SMP 16  & $4.844\pm 0.004$ & $3.47\pm 0.04$ & $1.2\pm 0.2$   & 5.5  \\
SMP 17  & $4.863\pm 0.004$ & $4.30\pm 0.22$ & $4.5\pm 1.5$   & 0.15 \\
SMP 18  &                  &                &                &      \\
SMP 19  & $5.227\pm 0.023$ & $3.78\pm 0.22$ & $2.3\pm 0.4$   & 0.83 \\
SMP 20  & $4.834$          & $3.75\pm 0.04$ & $1.0\pm 0.1$   & 9.9  \\
SMP 21  & $5.195\pm 0.034$ & $3.28\pm 0.2$  & $1.0\pm 0.0$   & 9.9  \\
SMP 22  & $5.326\pm 0.014$ & $3.80\pm 0.13$ & $1.5^{**}$     & 2.7  \\
SMP 23  & $4.903\pm 0.004$ & $3.09\pm 0.09$ & $0.9$          & 14.8 \\
SMP 24  &                  &                &                &      \\
SMP 25  & $5.256\pm 0.057$ & $3.33\pm 0.30$ & $1.3\pm 0.1$   & 5.6  \\
SMP 26  & $5.289\pm 0.027$ & $3.19\pm 0.24$ & $2.2\pm 0.2$   & 0.95 \\
SMP 27  & $4.869$          & $3.73\pm 0.05$ & $1.3\pm 0.1$   & 5.3  \\
SMP 28  & $5.293\pm 0.016$ & $3.37\pm 0.10$ & $1.5\pm 0.0$   & 2.7  \\
SMP 32  &                  &                &                &      \\
SMP 34  & $5.008$          & $2.56\pm 0.05$ & $0.89^*$       & 15.0 \\
MGPN 1    & $5.091$          & $4.36\pm 0.17$ & $5.0\pm 1.5$   & 0.09 \\
MGPN 2    & $4.998$          & $5.13\pm 0.05$ &                &      \\
MGPN 3    & $5.088$          & $2.76\pm 0.05$ & $0.9^{**}$     & 15.0 \\
MGPN 5    & $5.503\pm 0.039$ & $2.71\pm 0.14$ &                &      \\
MGPN 6    & $5.530$          & $2.93\pm 0.13$ &                &      \\
MGPN 7    & $5.156\pm 0.091$ & $3.55\pm 0.34$ & $1.3\pm 0.3^*$ & 5.2  \\
MGPN 8    & $5.146$          & $2.98\pm 0.01$ & $1.4\pm 0.0$   & 3.7  \\
MGPN 9    & $4.841$          &                &                &      \\
MGPN 10   & $5.530$          & $3.26\pm 0.10$ &                &      \\
MGPN 11   & $3.88\pm 0.38$ & $2.0\pm 1.0$   & 1.1   \\
MGPN 12   & $5.297$          &                &                &      \\
MGPN 13   & $5.530\pm 0.076$ & $3.37\pm 0.26$ &                &      \\
Ma 01   &                  &                &                &      \\
Ma 02   &                  &                &                &      \\
{[M95] 8}     &                  &                &                &      \\
{[M95] 9}     &                  &                &                &      \\
\hline
\end{tabular}
\begin{list}{}{}
\item[$^{a}$] * upper limit, ** lower limit
\end{list}
\end{table*}

%
%--------------------------------------------------------------------------------

%
% 
%--------------------------------------------------------------------------------

\section{The age-metallicity relation}

\begin{figure}
   \centering
   \includegraphics[angle=-90, width = 8cm]{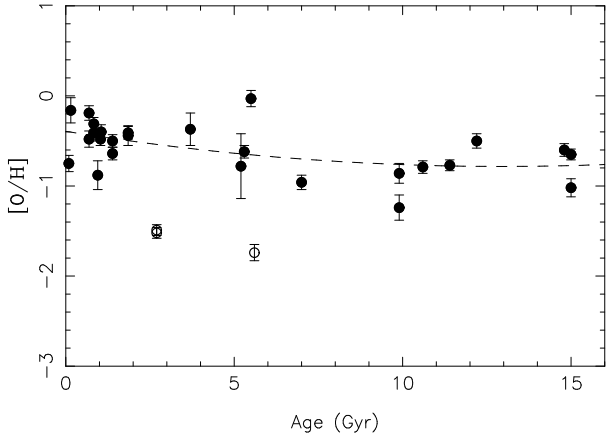}
   \caption{Oxygen abundances as a function of age. A second order polynomial
    fit is also included (dashed line). Objects plotted as empty circles are
    not included in the fit.}
   \label{amroh}
   \end{figure}

As a first approximation, O, Ne, S, and Ar measured in PNe can be used as tracers
of interstellar abundances at the time the progenitor stars were born. Since these
elements are generally well correlated with each other, as we have seen in
Fig.~\ref{temperature}, in the following we will consider oxygen abundances as
representative, since the oxygen lines are very bright and the abundances are 
well determined. Fig.~\ref{amroh} gives the derived age-metallicity relation in the form 
of oxygen abundances as a function of age. The abundances are defined as [O/H] = 
log(O/H) -- log(O/H)$_\odot$, as usual, and the solar abundance was taken as 
log(O/H)$_\odot$ + 12 = 8.70 (see for example Allende-Prieto et al. \cite{allende} and 
Asplund et al. \cite{asplund}).  Error bars are included for the oxygen abundances according 
to the values given in Table~\ref{elemental}. Concerning the stellar ages, it is difficult to 
estimate meaningful uncertainties, but based on the adopted isochrones, a typical error of 
about $\Delta t \simeq 0.5\,$Gyr can be estimated for ages $t \leq 4\,$Gyr, and
of $\Delta t \simeq 2\,$Gyr for older objects.  

%
%--------------------------------------------------------------------------------
%--------------------------------------------------------------------------------

The three young objects in Fig.~\ref{amroh}  with low oxygen abundances
and represented as empty circles are SMP22, SMP25, and SMP28. These nebulae have extremely
large N/O ratios, log(N/O) $\simeq 0.8$, compared to the other objects. Since their
nitrogen abundance relative to hydrogen is normal, log(N/H)+12 $\simeq 8$, their
oxygen abundance is strongly depressed, probably owing to ON cycling in the progenitor
star. Therefore these objects should not be included in the determination of the
age-metallicity relation. 
 
\begin{figure}
   \centering
   \includegraphics[angle=-90, width = 8cm]{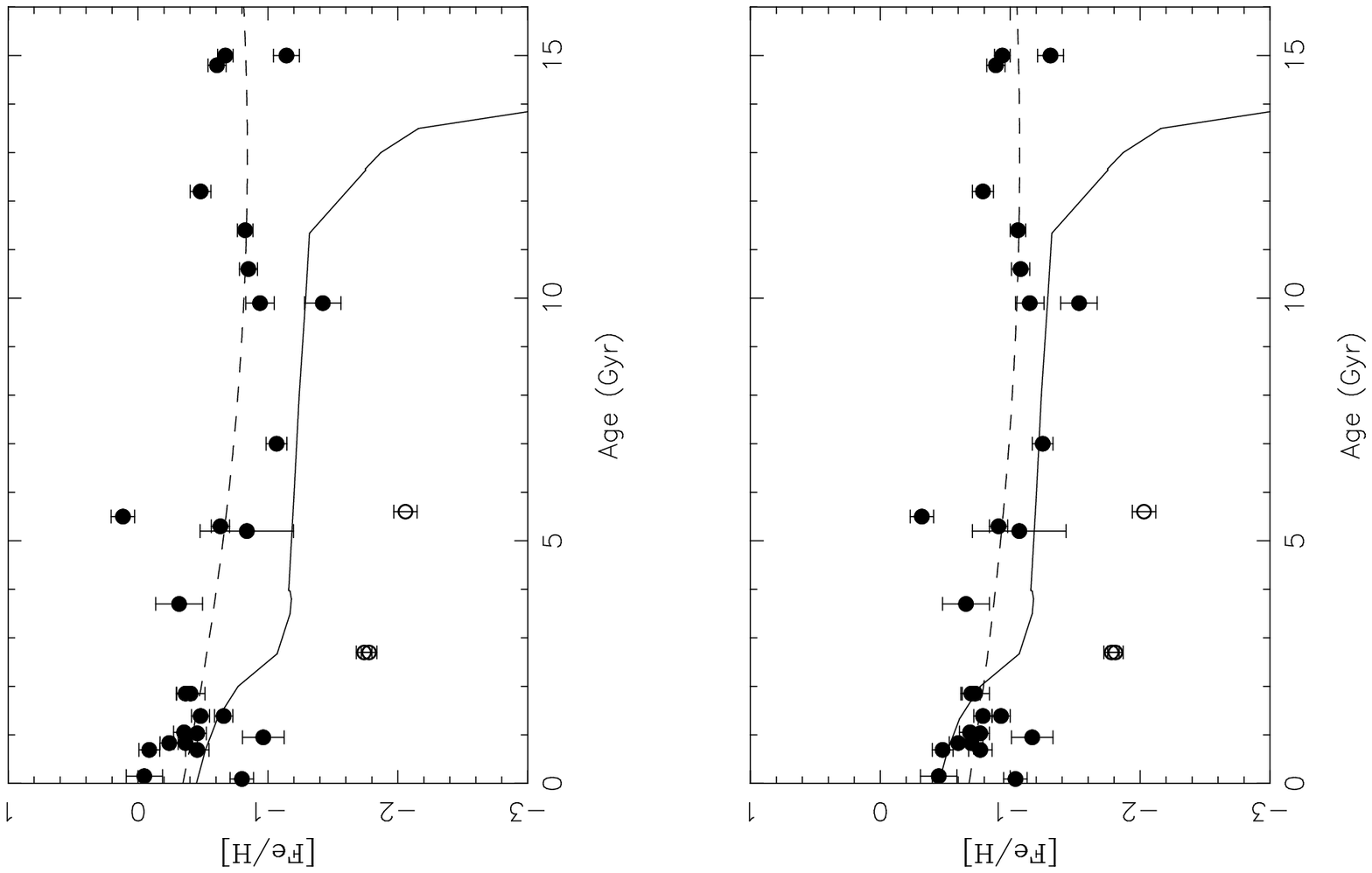}
   \caption{[Fe/H] metallicities as a function of age. The solid line shows the
    results of theoretical models for the SMC by Pagel \& Tautvai\v siene 
    (\cite{pagel}). The data points show  (a) the converted [Fe/H] metallicities
    using  the [O/Fe] $\times$ [Fe/H] relationship from Matteucci (\cite{matteucci}). 
   (b)  the converted [Fe/H] metallicities obtained by calibrating the [O/H] $\times$ 
   [Fe/H] relationship matching the abundances of the youngest objects in our sample 
   and in the models by Pagel \& Tautvai\v siene (\cite{pagel}).}
   \label{amrfeh}
   \end{figure}

In order to compare our derived age-metallicity relation with stellar data, it is interesting
to convert the [O/H] abundances into [Fe/H] metallicities. Iron cannot be accurately
determined from PNe data, in view of the faintness of the Fe lines and the possibility
of grain condensation, but average [O/Fe] $\times$ [Fe/H] relationships can be obtained
from stellar data or theoretical models. In the range $-2.0 < {\rm [Fe/H]} < 0.0$ 
this relationship is approximately linear, so that  an [O/H] to [Fe/H] conversion formula 
can be written as

   \begin{equation}
    {\rm [Fe/H]} \simeq a + b\, {\rm [O/H]}  \ .
   \label{fehoh}
   \end{equation}

\noindent
where $a$ and $b$ are constants in this metallicity range. As an example, the
theoretical SMC model by Russell \& Dopita (\cite{russell}) leads to the coefficients 
$a \simeq 0.45$ and $b \simeq 1.41$. Alternatively, using the theoretical [O/Fe] 
$\times$ [Fe/H] relation for the Magellanic Clouds given by Matteucci (\cite{matteucci}, 
fig. 6.3), one obtains $a \simeq 0.16$ and $b \simeq 1.27$. Also, using the approximate
[O/H] $\times$ age relation for the SMC planetary nebulae, and an [Fe/H] $\times$ age
relation obtained from the SMC cluster data of Da Costa \& Hatzidimitriou (\cite{dacosta1}), 
de Freitas Pacheco et al. (\cite{pacheco}), Piatti et al. (\cite{piatti1}, \cite{piatti2}),
and Da Costa (\cite{dacosta2}), we have $a \simeq 0.15$ and $b \simeq
1.28$, very similar to the previous set of coefficients. The corresponding [O/Fe] 
$\times$ [Fe/H] relation from these sources present a good fit with stellar data
from several sources, as can be seen for example from Hill et al. (\cite{hill}).  
Fig.~\ref{amrfeh}a shows the [Fe/H] $\times$ age relationship obtained  using 
Eq.~\ref{fehoh} with the constants derived from  the [O/Fe] $\times$ [Fe/H] relation 
for the SMC from Matteucci (\cite{matteucci}). The other sets of
coefficients mentioned above would lead to a figure very similar to 
Fig.~\ref{amrfeh}a.  Since we would like to compare this
relationship with some work done previously to the derivation of the new solar oxygen
abundances, we have used in the calibration of Fig.~\ref{amrfeh} the older solar
abundances, namely log (O/H)$_\odot$ + 12 = 8.93 (cf. Anders \& Grevesse \cite{anders}).
A similar plot based on preliminary results was presented by Idiart et al. (\cite{icm2005}).

%
%--------------------------------------------------------------------------------
%--------------------------------------------------------------------------------

The results of Fig.~\ref{amroh} or Fig.~\ref{amrfeh} can be conservatively interpreted 
as a mild decrease of the metallicity with age, as shown by the dashed line, which is a 
least squares fit to  the data adopting a second order polynomial, and not including the 
objects represented as empty circles. However, a closer look suggests that there is a 
sharper increase in the abundances in the last 2 Gyr, whereas for ages larger than 3 Gyr 
the abundances do not change appreciably, which is consistent with a recent burst of star 
formation. For illustration purposes, Fig.~\ref{amrfeh} includes the theoretical 
age-metallicity predicted by the model of Pagel \& Tautsvai\v siene (\cite{pagel}),  
which clearly indicates a burst in the last 2 to 3 Gyr. It is not our purpose here to 
fit the observed age-metallicity relation to theoretical models, but it should be 
mentioned that the agreement between our results and the models by Pagel \& Tautvai\v siene
(\cite{pagel}) can be improved by calibrating the oxygen abundances directly with the
model input. In this case, we could adopt [Fe/H] $\simeq -0.45$ for the youngest objects 
as in Pagel \& Tautvai\v siene (\cite{pagel}), which corresponds to log (O/H) + 12 $\simeq 
8.54$ in our sample, or [O/H] $\simeq -0.39$ [using log (O/H)$_\odot$ + 12 = 8.93], so that 
we would have [O/Fe] $\simeq 0.06$ and the metallicity would be given by [Fe/H] $\simeq$ 
[O/H] -- 0.06, instead of Eq.\ref{fehoh}. Fig.~\ref{amrfeh}b shows the corresponding 
age-metallicity relation, which is better adjusted to the data, especially near the burst 
region at $t < 3\,$Gyr.

The presence of bursts in the Magellanic Clouds has already been suggested in some previous 
work. For instance, in the case of the LMC some signs of star formation bursts can be seen 
in the age-metallicity relation derived from HST observations of PNe by Dopita et al.
({\cite{dopita1}, fig. 5) and Dopita (\cite{dopita2}). In this case, a strong burst of 
star formation was suggested between 1--3 Gyr ago, which increased the metallicity
by a factor larger than two. Stellar cluster data also support these conclusions, as 
can be seen from van den Bergh (\cite{vandenbergh}) and Girardi et al. (\cite{girardi}).
More generally, recent models of Blue Compact Galaxies and dwarf spheroidal
galaxies in the Local Group were developed by Lanfranchi \& Matteucci 
(\cite{lanfranchi}) who suggested that the former are characterized by several short 
bursts of star formation separated by long quiescent periods, while for the latter
one or two long bursts and efficient stellar winds would be sufficient.

For the SMC, the age distribution of clusters seems to be more homogeneous (cf. van den
Bergh \cite{vandenbergh}), a result that is supported by the essentially monotonic
age-metallicity relation derived by Dolphin et al. (\cite{dolphin}) on the basis
of SMC clusters and field stars. More recently, Harris \& Zaritsky (\cite{harris})
determined the global star formation history (SFH) and the age-metallicity relation
for the SMC based on their photometric survey. In this case, a distinct increase
in the global metallicity occurred in the last 3 Gyr, rising the metallicity from
[Fe/H] $\simeq -0.9$ to [Fe/H] $\simeq -0.4$, which is consistent with our results,
taking into account the uncertainties in the metallicity determination. As discussed by
Harris \& Zaritsky (\cite{harris}), that period coincided approximately with a past
perigalactic passage by the SMC relative to the Milky Way, which may have originated
the observed burst. These results are largely consistent with cluster
data by Da Costa \& Hatzidimitriou (\cite{dacosta1}) and Piatti et al. (\cite{piatti1},
\cite{piatti2}). Also, in a recent work, Kayser et al. ({\cite{kayser}) obtained
VLT spectroscopy and HST photometry for a large sample of SMC clusters. Using a 
[Fe/H] calibration and adopted ages, they derived an age-metallicity relation and
found a flat plateau between 2 and 4 Gyr approximately, and a steep rise  towards
the younger end, which is in excellent agreement with our results shown in
Fig.~\ref{amrfeh}. In another recent work on the SFH of the SMC based on cluster
data, No\"el et al. (\cite{noel}) obtained average metallicities at several age
intervals for cluster ages $t < 13\,$Gyr and found that the mean metallicities
of the stars formed in the considered SMC field were about [Fe/H] $\simeq -1.3$
for $t > 4\,$Gyr, increasing steadily to the present gas-phase abundance
of about [Fe/H] $\simeq -0.5$. Although these results are still preliminary,
and represent a limited region of the SMC, it is interesting to note that
a very good agreement is obtained with our present results.

\section{Conclusions}

In this paper we have presented new abundance data on planetary nebulae
in the SMC, in order to study the chemical evolution of this galaxy. Spectral
line fluxes are given for a sample of 36 nebulae, and the chemical composition
is derived for a larger sample of 44 objects. The physical properties of
most of the PNe progenitor stars are derived, including the effective 
temperatures, luminosities, masses, and ages. As a result, an age-metallicity
relation was obtained for the SMC, which shows a clear indication of a 
star formation burst that occurred 2--3 Gyr ago. By the use of a calibration
between the Fe and O abundances, this relation can be compared with similar
relations obtained recently, acting as an important constraint to chemical
evolution models for the SMC.

\begin{acknowledgements}
      We thank Dr. Laura Magrini for some interesting suggestions. This work was 
      partially supported by FAPESP and CNPq.
\end{acknowledgements}


\begin{thebibliography}{}


    \bibitem[2001]{allende} Allende-Prieto, C., Lambert, D. L., \& Asplund, M.
       2001, ApJ, 556, L63

    \bibitem[1984]{aller2} Aller, L. H. 1984, Physics of thermal gaseous nebulae
      (Dordrecht: Reidel)

    \bibitem[1981]{aller} Aller, L. H., Keyes, C. D., Ross, J. E., \& Omara, B. J. 1981, 
         MNRAS, 194, 613

    \bibitem[1989]{anders} Anders, E. \& Grevesse, N. 1989, Geochim. Cosmochim. Acta
       53, 197

    \bibitem[2004]{asplund} Asplund, M., Grevesse, N., Sauval, A. J., Allende-Prieto, C.,
    \& Kiselman, D. 2004, A\&A, 417, 751

    \bibitem[1992]{bertelli} Bertelli, G., Mateo, M., Chiosi, C., \& Bressan, A. 1992, 
         ApJ, 388, 400

    \bibitem[1989]{boronson} Boroson, T. A. \& Liebert, J. 1989, ApJ, 339, 844 

    \bibitem[1989]{cardelli} Cardelli, J. A., Clayton, G. C., \& Mathis, J. S. 1989, 
        ApJ, 345, 245

    \bibitem[1994]{cazetta} Cazetta, J. O. \& Maciel, W. J. 1994, A\&A, 290, 936
        ApJ, 345, 245

    \bibitem[1994]{chiappini} Chiappini, C., \& Maciel, W. J. 1994, A\&A, 288, 921

    \bibitem[2000]{costa} Costa, R. D. D., de Freitas Pacheco, J. A., \& Idiart, T. P. 
          2000, A\&AS, 145, 467

    \bibitem[2002]{dacosta2} Da Costa, G. S. 2002, IAU Symposium 207, Extragalactic
      Star Clusters, ed. D. Geisler, E. K. Grebel, \& D. Minniti, (San Francisco: ASP), 83

    \bibitem[1998]{dacosta1} Da Costa, G. S., Hatzidimitriou, D. 1998, AJ, 115, 1934

    \bibitem[1998]{pacheco} de Freitas Pacheco, J. A., Barbuy, B., \& Idiart, T. P. 
          1998, A\&A, 332, 19

    \bibitem[1989]{dennefeld} Dennefeld, M. 1989, Recent developments of Magellanic
        Cloud research (Meudon: Observatoire de Paris), ed. K. S. de Boer, 
        F. Spite, \& G. Stasi\'nska, 107

    \bibitem[2001]{dolphin} Dolphin, A. E., Walker, A. R., Hodge, P. W., et al. 2001, 
         ApJ, 562, 303 

    \bibitem[1999]{dopita2} Dopita, M. A. 1999, New views of the Magellanic Clouds, IAU
      Symp. 190, ed. Y. -H. Chu, N. B. Suntzeff, J. E. Hesser, \& D. A. Bohlander
      (San Francisco: ASP), 332

    \bibitem[1997]{dopita1} Dopita, M. A., Vassiliadis, E., Wood, P. R., et al. 1997, 
       ApJ, 474, 188

    \bibitem[1977]{dufour} Dufour, R. J. \& Killen, R. M. 1977, ApJ, 211, 68

    \bibitem[2004]{escudero} Escudero, A. V., Costa, R. D. D., \& Maciel, W. J. 2004, A\&A, 
        414, 211

    \bibitem[1987]{feast} Feast, M. W. \& Walker, A. R. 1987, ARA\&A, 25, 345

    \bibitem[1995]{girardi} Girardi, L., Chiosi, C., Bertelli, G., \& Bressan, A.
        1995, A\&A, 298, 87

    \bibitem[2003]{harries} Harries, T., J., Hilditch, R. W., \& Howarth, I. D. 2003, 
        MNRAS, 339, 157

    \bibitem[2004]{harris} Harris, J., \& Zaritsky, D. 2004, AJ, 127, 1531

    \bibitem[2004]{henry} Henry, R. B. C., Kwitter, K. B., \& Balick, B. 2004, AJ, 127, 2284

    \bibitem[1997]{hill} Hill, V., Barbuy, B., \& Spite, M. 1997, A\&A, 323, 461

    \bibitem[2005]{icm2005} Idiart, T. P., Costa, R. D. D., \& Maciel, W. J. 2005, 
        Planetary Nebulae as astronomical tools, ed. R. Szczerba, G. Stasi\'nska,
        \& S. K. G\'orny (Melville, NY: Am. Inst. of Physics), 261

    \bibitem[1989]{kaler} Kaler, J. B. \& Jacoby, G. H.  1989, ApJ, 345, 871

    \bibitem[2007]{kayser} Kayser, A., Grebel, E. K., Harbeck, D. R., Cole, A. A.,
      Koch, A., Gallagher, J. S., \& Da Costa, G. S. 2007, Globular clusters - guides
      to galaxies, ed. T. Richtler \& S. S. Larsen, (Heidelberg: Spring) (in press)
      (astro-ph/0607047)

    \bibitem[2006]{keller} Keller, S. C., \& Wood, P. R. 2006, ApJ, 642, 834

    \bibitem[1995]{kingdon} Kingdon, J., \& Ferland, G. J. 1995, ApJ, 442, 714

    \bibitem[1994]{kingsburgh} Kingsburgh, R. L., \& Barlow, M. J. 1994, MNRAS, 271, 257

    \bibitem[2003]{lanfranchi} Lanfranchi, G., \& Matteucci, F. 2003, MNRAS, 345, 71

    \bibitem[2006]{leisy2} Leisy, P. 2006, private communication

    \bibitem[1996]{leisy1} Leisy, P. \& Dennefeld, M.  1996, A\&AS, 116, 95

    \bibitem[2006]{leisy3} Leisy, P. \& Dennefeld, M.  2006, A\&A, 456, 451

    \bibitem[2006]{mci2006} Maciel, W. J., Costa, R. D. D., \& Idiart, T. P. 2006, 
       Planetary Nebulae beyond the Milky Way, ed. L. Stanghellini, J. R. Walsh, \& 
       N. G. Douglas (Heidelberg: Springer), 209

    \bibitem[2007]{mqc2007} Maciel, W. J., Quireza, C., \& Costa, R. D. D. 2007, A\&A, 463, L13

    \bibitem[2000]{matteucci} Matteucci, F. 2000, The chemical evolution of the
       Galaxy (Dordrecht: Kluwer)

    \bibitem[1991a]{meatheringham1} Meatheringham, S. J., \& Dopita, M. A. 1991a, 
         ApJS, 75, 407

    \bibitem[1991b]{meatheringham2} Meatheringham, S. J., \& Dopita, M. A. 1991b, 
        ApJS, 76, 1085

    \bibitem[1988]{meatheringham3} Meatheringham, S. J., Dopita, M. A., \& Morgan, 
         D.H. 1988, ApJ, 329, 166

    \bibitem[1995]{meyssonier} Meyssonnier, N. 1995, A\&AS, 110, 545


    \bibitem[1988]{monk} Monk, D. J., Barlow, M. J., \& Clegg, R. E. S. 1988, MNRAS, 
        234, 583

    \bibitem[2007]{noel} No\"el, N. E. D., Gallart, C., Aparicio, A., Hidalgo, S. L.,
        Carrera, R., Costa, E., \& M\'endez, R. A. 2007, Stellar populations as 
        building blocks of galaxies, IAU Symp. 241, ed. A. Vazdekis, \& R. Peletier,
        (San Francisco: ASP) (in press)

    \bibitem[1996]{olszewski} Olszewski, E.W., Suntzeff, N.B., \& Mateo, M. 1996, 
        ARA\&A, 34, 511

    \bibitem[1976]{osmer} Osmer, P. S. 1976, ApJ, 203, 352

    \bibitem[1989]{osterbrock} Osterbrock, D. E. 1989, Astrophysics of Gaseous Nebulae 
        and Active Galactic Nuclei (Mill Valley: University Science Books) 

    \bibitem[1998]{pagel} Pagel, B. E. J.. \& Tautvai\v siene, G. 1998, MNRAS, 299, 535 

    \bibitem[1978]{peimbert} Peimbert, M. 1978, Planetary Nebulae, IAU Symp. 76,  
        ed. Y. Terzian (Dordrecht: Reidel), 215

    \bibitem[1991]{daniel} P\'equignot, D., Petitjean, P., \& Boisson, C. 1991, A\&A, 251, 680

    \bibitem[2001]{piatti1} Piatti, A. E., Santos, J. F. C., Clari\'a, J. J., Bica, E.,
        Sarajedini, A., \& Geisler, D. 2001, MNRAS, 325, 792

    \bibitem[2005]{piatti2} Piatti, A. E., Sarajedini, A., Geisler, D., Seguel, J., 
        \& Clark, D. 2005, MNRAS, 358, 1215

    \bibitem[2006]{richer} Richer, M. G., \& McCall, M. L. 2006, Planetary Nebulae 
       beyond the Milky Way, ed. L. Stanghellini, J. R. Walsh, \& N. G. Douglas 
       (Heidelberg: Springer), 220

    \bibitem[1994]{rola} Rola, C. \& Stasi\'nska, G. 1994, A\&A, 282, 199 

    \bibitem[1992]{russell} Russell, S. C., \& Dopita, M. A. 1992, ApJ, 384, 508

    \bibitem[2003]{stanghellini} Stanghellini, L., Shaw, R. A., Balick, B., Mutchler, M., 
        Blades, J. C., \& Villaver, E. 2003, ApJ, 596, 997

    \bibitem[1998]{stasinska} Stasi\'nska, G., Richer, M. G., \& McCall, M. L. 1998, A\&A, 336, 667

    \bibitem[1991]{vandenbergh} van den Bergh, S. 1991, ApJ, 369, 1

    \bibitem[1992]{vassiliadis} Vassiliadis, E., Dopita, M. A., Morgan, D. H.,  \& 
         Bell, J. F. 1992, ApJS, 83, 87

    \bibitem[1994]{vassiliadis2} Vassiliadis, E.,  \&  Wood, P. R. 1992, ApJS, 92, 125

%    \bibitem[2004]{villaver} Villaver, E., Stanghellini, L., \& Shaw, R. A. 2004, ApJ, 
%        614, 716 

    \bibitem[1976]{webster} Webster, B. L. 1976, MNRAS, 174, 513

    \bibitem[1987]{wood} Wood, P. R., Meatheringham, S. J., Dopita, M. A., \& Morgan, D. H. 
        1987, ApJ, 320, 178


\end{thebibliography}
\end{document}